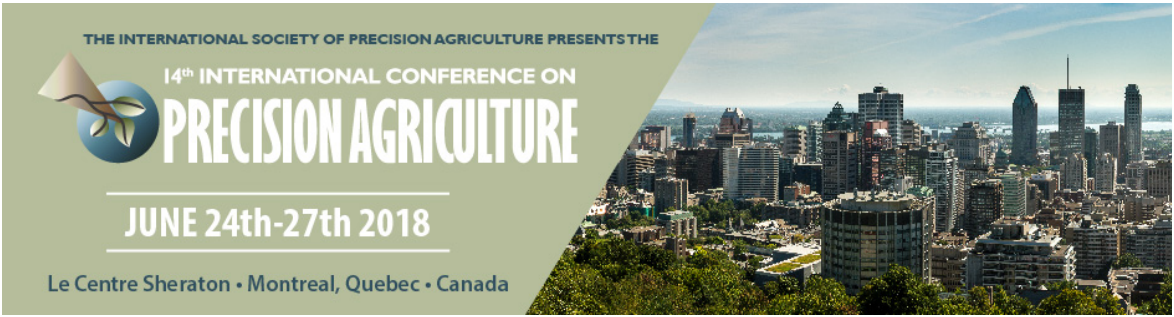

# ANALYZING TRENDS FOR AGRICULTURAL DECISION SUPPORT SYSTEM USING TWITTER DATA

↑Sneha Jha, *Dharmendra Saraswat, °Mark Daniel Ward

↑Graduate Student, Agricultural and Biological Engineering, Purdue University, Indiana

*Associate Professor, Agricultural and Biological Engineering, Purdue University, Indiana

°Associate Professor, Department of Statistics, Purdue University, Indiana

***Abstract.*** *The trends and reactions of the general public towards global events can be analyzed using data from social platforms, including Twitter. The number of tweets has been reported to help detect variations in communication traffic within subsets like countries, age groups and industries. Similarly, publicly accessible data and (in particular) data from social media about agricultural issues provide a great opportunity for obtaining instantaneous snapshots of farmers' opinions and a method to track changes in opinion through temporal analysis. In this paper we hypothesize that the presence of keywords like precision agriculture, digital agriculture, Internet of Things (IoT), BigData, remote sensing, GPS, etc., in tweets could serve as an indicator of discussions centered around interest in modern farming practices. We extracted relevant tweets using keywords such as IoT, BigData and Geographical Information System(GIS), and then analyzed their geographical origin and frequency of their mention. We analyzed the Twitter data for the period of $1^{st}$-$11^{th}$ January, 2018 to understand these trends and the factors affecting them. These factors, such as special events, projects, biogeography, etc., were further analyzed using tweet sources and trending hashtags from the database. The regions with the highest interest in the keywords were United States, Egypt, Brazil, Japan and China. A comparison of frequency of keywords revealed IoT as the most tweeted word (77.6%) in the downloaded data. The most used language was English followed by Spanish, Japanese and French. Periodical tweets on #IoT from an account handled by IoT project on Twitter and Seminars on IoT in January in Santa Catarina(Brazil) were found to be the underlying factors for the observed trends.*

## Introduction

Social media platforms have evolved from being an online individual communication media in its early days (Orkut, Myspace, Facebook) to their present forms (live streaming, blogging and customizable aggregators of user preference media, such as Twitter, Instagram, etc.). This evolution has made Twitter a favorable platform of communication for varied groups ranging from business, education, press and public. Twitter in the modern world acts as platform for accumulation of news, media, and social information. It acts as a strong social influencer because it has been filtered through trusted sources i.e. friends, family, followers *(Boulianne 2015*; *Hawn, 2009)*. The use of data from social media like Twitter presents a higher association to the reaction of users towards any kind of news and information due to the curated content. Acquisition of publicly available data on Twitter is through various options, such as high-cost firehose (access to 100% data), decahose (low-cost, randomly sampled, 10% of the data), or real time streaming, which allows 1% of the daily data to be downloaded. The method used to download data depends on the extent and the type of analysis required. Even if real time streaming data is freely available, it does not have any fewer attributes or any missing information, as compared to firehose or decahose, i.e., paid data methods (*Morstatter et al., 2012)*. The Twitter data has been used to create Healthcare models in real time by using cloud storage and then creating a word cloud from downloaded data, to indicate the increase in vaccination program implementation in recent years (*Saini et al., 2018)*. Analyzing Twitter data has four main stages: data collection, data loading, Tweet classification and storage and analysis (*Khan et al., 2018).* Twitter data analysis can be both quantitative, like the healthcare vaccination count, or be based on reaction or sentiment. Quantitative analysis includes relating the frequency of tweets to a user, time, location, or language, to infer useful information or trends (*Sinnenberg et al., 2017)*. The analysis of Twitter data to estimate the sentiment of the words and classify them into emotions that are either positive, negative, or neutral; this is part of sentiment analysis.

Data use is continually increasing in agricultural sector due to an increased utilization of a variety of sensors and IoT devices in modern agricultural machines and on farms. The response of farmers towards enhanced modernization of their workplace is important for industry and research institutions who are engaged in developing and introducing new machines for efficiently and effectively performing farming operations. Typically, a famer's feedback is gathered through customized surveys but increased use of social networking sites makes it an attractive option for obtaining feedback.

Integration of Twitter into decision-making process by industry and research institutions could lead to developing an integral platform for sharing expert knowledge about precision agricultural solutions and gathering feedback from users. To create such a system would require a quantitative and qualitative analysis of the generated data.

This study aims to analyze the trends in precision agricultural solutions by a quantitative analysis of relevant tweets. This paper focuses on using BigData, IoT and GIS as the precision agricultural keywords to filter tweets for a quantitative analysis of their origin, language and frequency.

## Materials and methods

This study is a quantitative analysis of precision agriculture related tweets to study the trends in new methods of farming in the society. This section is used to describe the methods used to obtain the tweets related to precision agriculture and their analysis.

To acquire the Twitter data, we first developed a list of keywords, such that the presence of these keywords would indicate that the tweets occurred in the context of precision agriculture. These keywords were selected after repetitive search of hashtags such as, PrecisionAg, precision agriculture, Agriculture, Digital Agriculture, DigitalAg, DecisioninAG, Agtech, AgricultureTechnology, trendsinAG, TechAg, BigdataAG, AGBigData, BigData, IOT, GIS, etc .

Among these keywords IOT, BigData and GIS were used as the filtering keywords in this study, due to their higher presence in Twitter search results and their overall popularity among precision agriculture technologies.

Filtering of keywords used a comparison method, which was case and character sensitive, so it was vital to have the words spelled exactly similar to the ones appearing in most of the Twitter searches. Then to download the tweets, we first made an active Twitter profile. This profile was required to create an app which is designed by Twitter to allow access to its freely available real time streaming tweet data using authentication keys and tokens. These keys and tokens are unique authentication keys and can be used only with one user profile.

The tweets were downloaded using Python APIs using Tweepy and Twitterscraper modules. To ensure continuous connection, a virtual environment was set up in the Purdue community cluster that was not affected by external system files and worked even when the local machine was shut down. This was done to ensure continuous download of tweets over a period of multiple days and weeks without broken data points.

Real time streaming data was filtered based on the following keywords in the virtual environment, downloaded into JSON file format and stored in the cluster for large data files. The downloaded tweets followed a predefined format called the tweet data dictionary, which used keys within parenthesis ("{} ") and values separated by commas. The information contained for each key was organized as per the structure shown in Table 1.

| Key | values | Data format |
| --- | --- | --- |
| **User_id** | User object (id, name, followers, friends, favorites, tweets, geo_enabled etc) | String of integers unique to each user, integers, boolean values |
| **Lang** | Machine detected language | String |
| **Coordinates** | Coordinates | geoJSON |
| **place** | Country_code, country , city, id | String of numbers or characters |
| **text** | text of tweets, emojis, links, images | UTF-8 text format |

**Table 1. Data structure in tweets**

*Quantitative analysis of the downloaded tweets:*

We created a statistical representation for the quantitative analysis of precision agriculture related keywords, to better understand the outreach of these words among the social media using population of the world. The data was downloaded during 1$^{st}$-11$^{th}$ January, 2018 and contained 3821 tweets. The keywords used for the quantitative analysis were the same as the keywords used to filter real time tweets, i.e., IoT, BigData and GIS, as these were the buzz words used in the context of precision agriculture on Twitter. Also the analysis of these new technologies in the modern methods of farming would be useful to analyze the trends prevalent in agriculture technological solutions. The attributes used for the analysis included language, place and frequency of keywords.  The matplotlib and pandas modules in Python were used for the visualization of as discussed below.

*Place:*

This key contains information about the place of origin of the tweet. This key was used to count the number of tweets originating from a country or location. The country and location of the tweets indicate the geographical features which influence the demand or feasibility of the PA solutions. The countries were represented by country codes, which were sorted, stored into matrices and then again counted and sorted in an increasing order to get the top five countries or locations.

*Text:*

The text key contains the actual text posted by the user and is complex in nature due to the types of characters used in the text which include emojis, white spaces, special characters, links and images. The text was processed before it was used for analysis. The extracted text from the data repository was first converted into lower case strings to facilitate searching of keywords by matching. The processed data was then used to count the number of tweets with keywords of interest (i.e. bigdata, iot and gis). A bar plot was created depicting the frequency of occurrence of keywords in the downloaded data.

This key also contains a string with the code for the language of the tweet. The language of the tweets is important because it is based on user's preference, instead of the country or location of origin. The language related data was used to plot the top five used languages in the Twitter downloaded data, similar to the one created for the location of tweets.

The obtained statistical results were then used to find the underlying event or cause which effected the trends. A manual search of the database was performed for determining source of the tweets, the main hashtags for events, and then an extensive search for those events was performed on Twitter. This effort led to finding the events or accounts which were responsible for increased Twitter data during the time of data acquisition. This was an important step in analyzing the effect of people and real world events on the flow of information on Twitter.

## Results and discussion

The trends in the tweets containing the keywords based on their origin, frequency and language are reported in this section. These trends were corroborated with Twitter searches and have been discussed in the interpretations.

*Analysis of tweets based on origin:*

A geographical location of the tweets not only suggest that the tweet originated from the location but also that the tweet was in the region of interest of the user in that country or location. This kind of information is used by companies to set up a target consumer for further technological development or present technology transfers.

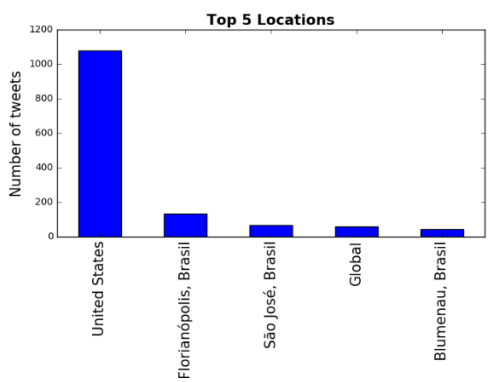 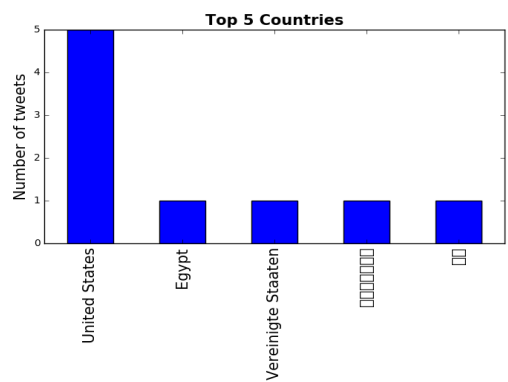

**Fig. 1(a): Top five locations of origin of the tweets in the PA repository**

**Fig. 1(b): Top five Countries of origin of the tweets in the PA repository**

*Analysis:* The graphs in Fig 1(a) and(b) show the top five locations and the top five countries the words were tweeted from in this database.

*Interpretation 1:* The values for locations and countries were not the same because the plotting with location used the 'city' value and plotting by country used the name of 'country' value to plot the bar graphs. User accounts with disabled geo-locations are named as Global in the downloaded data.

*Interpretation 2:* Some cities might have more tweet traffic during the data acquisition period yet that was not reflected for the whole country. Cities like Sao jose, Florianopolis, Blumenau (cities of Santa Catarina, Brazil) showed up on the highest ranking locations in the database (Fig. 1(a)) because of an IOT meetup in Florianopolis on 25th January, 2018 and twitter account @diomiciof5min, of SSP(Shared Storage Protocol) monitoring cameras, broadcasting data every 15 minutes on Twitter with #IoT. It seemed a most plausible reason for high traffic of tweets from the mentioned cities in January. On contrast, an analysis of the bar graph on country basis (Fig. 1(b)) we do not see Brazil showing up among the top five countries. Thus external events were a major cause for sudden increase in keyword frequency for a particular location but when considering the overall country data the local events did not cause much impact.

*Interpretation 3*: The third highest tweet was depicted from a country named Vereinigte Staaten which actually is United States but written in German. So the third highest number of tweets was by the people who were using German as their language but were located in the United States.

*Analysis based on frequency of precision Ag solution keywords*

Frequency analysis of the keywords in the downloaded database helped to identify those technologies that were more popular among the Twitter users. It could be used by companies as a surrogate measure of success of the product, and a temporal track of similar data over time may be used to gauge user interests during future offerings by agricultural industry.

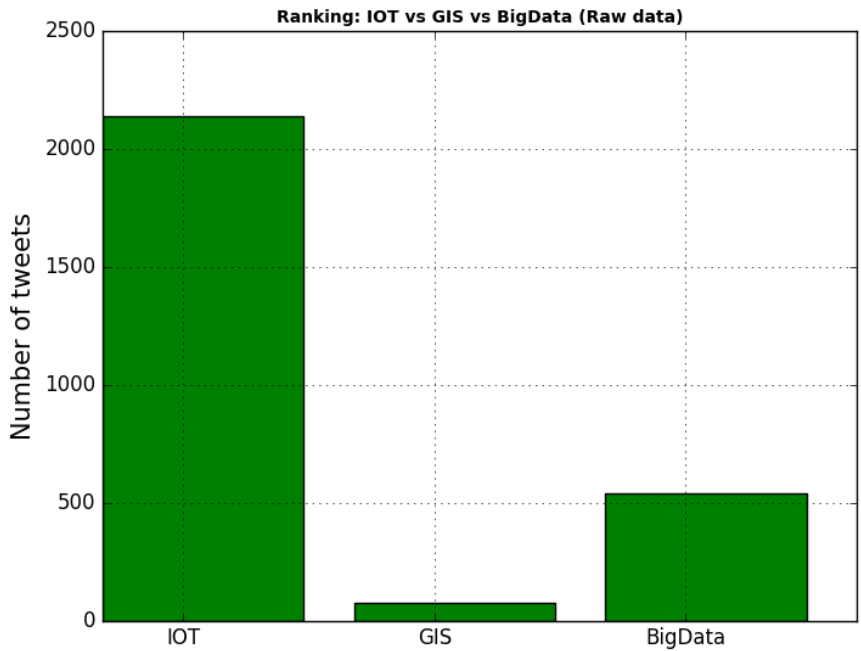

Fig 2: The graph shows the comparison of keyword frequency in the PA repository

*Analysis:* It is clear from Fig.3 that IoT was the most tweeted word followed by BigData and GIS. Based on the analysis of the database, it was found that 77.63% of the users mentioned IoT, 19.68% BigData, and 0.26% GIS in their tweets

*Interpretation 1:* IoT and BigData are generic technical terms that are not restricted to be used in agriculture only, instead they could also be used in the context of environment, health care, statistics etc. sectors. These words represent more universally integrated technological solutions and a contextual analysis is required for finding association with a sector in particular.

*Analysis based on the language of the tweet*

Tweets are usually written in the language used by the target audience so as to have maximum impact. Thus, analysis of the top languages of the tweets would exhibit a relation between marketing of these technologies on social media and usage or awareness of its target audience.

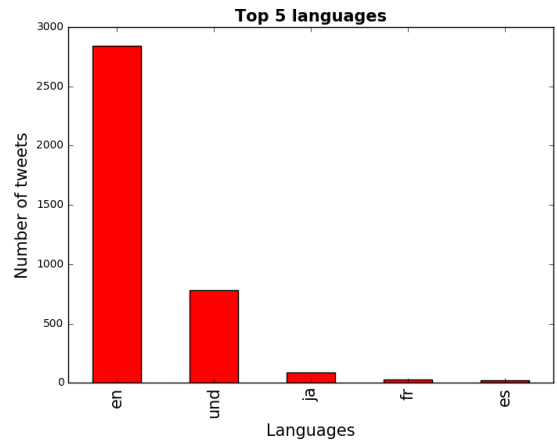

**Fig 3: The graph shows the top five languages of the tweets in the PA repository**

*Analysis:* Fig.3 shows that the top four languages found in the database were English, Undefined, Spanish and Japanese.

*Interpretation 1:* English being the most used language, suggests that the highest number of tweets are from English speaking countries.

*Interpretation 2:* The undefined represents the tweets which did not have the information about the language of the tweets.

*Interpretation 3:* Japan being an ardent user of technologies, Japanese showed up as the third highest tweeted language. Compared to English and Spanish, Japanese is spoken only by those living in Japan and a relatively small Japanese population living outside of Japan. A high number of tweets in Japanese language concerning these keywords are indicative of interests in new technologies in Japan.

## Conclusion

The analysis and interpretation of the Twitter data provided an insight to better understand interest in modern agricultural technological related keywords among public based on their location, tweeting pattern and the languages used. The study revealed that merely determining location of tweet is not enough but an accompanying causative analysis is needed to understand the context. In the long run, it is expected that the analysis and inferences drawn from Twitter data could become modern survey tools for industry and research institutions and help them make better market-related decisions.

# References

Sinnenberg, L., Buttenheim, A. M., Padrez, K., Mancheno, C., Ungar, L., & Merchant, R. M. (2017). Twitter as a tool for health research: a systematic review. American journal of public health, 107(1), e1-e8.

Morstatter, F., Pfeffer, J., Liu, H., & Carley, K. M. (2013, July). Is the Sample Good Enough? Comparing Data from Twitter's Streaming API with Twitter's Firehose. In ICWSM.

Boulianne, S. (2015). Social media use and participation: A meta-analysis of current research. Information, Communication & Society, 18(5), 524-538.

Khan, I., Naqvi, S. K., Alam, M., & Rizvi, S. N. A. (2018). A Framework for Twitter Data Analysis. In Big Data Analytics (pp. 297-303). Springer, Singapore.

Hawn, C. (2009). Take two aspirin and tweet me in the morning: how Twitter, Facebook, and other social media are reshaping health care. Health affairs, 28(2), 361-368.